\documentclass{jkas}

\usepackage{amsmath}

\def\year{2015} 
\def\month{February} 
\def\volume{48} 
\def\issue{0} 
\def\beginpage{1} 
\def\endpage{7} 
\def\received{---} 
\def\accepted{---} 
\date{Received \received; accepted \accepted}

\def\apj{{ ApJ}}
\def\apjl{{ ApJL}}
\def\apjs{{ ApJS}}

\def\jcap{{ J.\ Cosmol.\ Astropart.\ Phys.}}

\def\mnras{{ MNRAS}}

\usepackage{flushend}

\title{Light-Cone Effect of Radiation Fields in Cosmological\\ Radiative Transfer
Simulations}

\author{Kyungjin Ahn}

\affil{Department of Earth Sciences, Chosun University, Dong-Gu, Gwangju 501-759, Korea; \email{kjahn@chosun.ac.kr}}

\def\corrauthor{
K. Ahn
}

\def\runningauthor{
Ahn
}

\def\runningtitle{
Light-Cone Effect in Cosmological Radiation Transfer
}

\def\keywords{
cosmology: diffuse radiation --- methods: numerical
}

\def\abstracttext{
We present a novel method to implement time-delayed propagation of radiation
fields in cosmological radiative transfer simulations.  Time-delayed propagation of radiation
fields requires construction of retarded-time fields by tracking the
location and lifetime of radiation sources along the corresponding light-cones.
Cosmological radiative transfer simulations have, until now, ignored 
this ``light-cone effect'' or implemented ray-tracing methods that are
computationally demanding.
We show that radiative transfer calculation of the time-delayed
fields  can be easily achieved in numerical
simulations when periodic boundary conditions are used, by calculating
the time-discretized retarded-time Green's function using the Fast Fourier Transform
(FFT) method and convolving it with the source distribution. We also
present a direct application of this method to the long-range radiation
field of Lyman-Werner band photons, which is important in the
high-redshift astrophysics with first stars.
}

\begin{document}
\jkashead

\section{Introduction\label{sec:Introduction}}

Cosmological radiative transfer simulations are used to study astrophysical
processes which occur on a large, cosmic scale. A notable example
of such processes is the process of cosmic reionization, in which
individual H II regions take up large volumes, of order ($\sim20$
comoving Mpc)$^{3}$ or larger, after $\sim50\%$ of all the baryons
are reionized by astrophysical radiation sources (\citealt{2005MNRAS.363.1031F,2006MNRAS.365..115F,Iliev2007,Zahn2007}).
The typical radiation sources long after the recombination epoch are
stars and quasars, which emit predominantly ultra-violet (UV) and
X-ray photons, respectively. The mean free path of hydrogen-ionizing
(H-ionizing) UV photons is simply the average size of H II regions,\footnote{Note that even after the intergalactic medium (IGM) is fully ionized,
which is believed to occur at $z\lesssim 7$, the mean free path of
UV photons is severely restricted to a few comoving Mpc. This is due
to the rapid increase of the number of Lyman limit systems, which
contain neutral hydrogen atoms inside and thus consume hydrogen-ionizing
UV photons (e.g. \citealt{McQuinn2011,Alvarez2012}).%
} while the mean free path of X-ray photons is much larger than that
of UV photons due to the relatively small optical depth (the scattering
cross section of UV photons is much larger than that of X-ray photons:
e.g. \citealt{Mesinger2013,Xu2014}). Therefore, the mean free path
of X-ray photons is truly cosmological, surpassing several hundred
Mpc when the rest-frame photon energy is larger than a few keV (\citealt{Xu2014}).
Another example of long mean free paths is the Lyman-Werner (LW) band photons. These
photons lie at the energy range of 11 -- 13.6 eV, and thus traverse
cosmological distances freely until they redshift into hydrogen Lyman
resonance lines and are scattered off neutral hydrogen atoms (\citealt{Haiman2000,Ahn2009}).
Practically all LW band photons are scattered multiple times and reprocessed
into lower-energy photons when they traverse the ``Lyman-Werner horizon''
$r_{{\rm LW}}\sim100\,[21/(1+z)]^{0.5}$ Mpc, where $z$ is the redshift
at which the photons are emitted (\citealt{Ahn2009}).

The light-cone effect, which denotes the effect related to fields
travelling at the speed of light and thus tracing past-time events
along the corresponding light cones, is inherent in any physical processes
on cosmic scales. Radiation fields and gravitational fields are the
obvious examples governed by the light-cone effect: both fields travel
at the speed of light (see \citealt{Hwang2008} showing that the gravitational
field travels at the speed of light in the form of the electric part
of the Weyl tensor). The relevant time or length scale over
which the light-cone effect matters can be estimated by an effective
time-scale $t_{{\rm lc}}\equiv f/(df/dt)$, where $f$ is any physical
quantity relevant to the problem of interest and $df/dt$ is the change
rate of $f$. For astrophysical problems, the time scale may be naively
estimated by the lifetime of radiation sources of interest. In contrast,
gravitational fields under non-relativistic motion (velocity $\ll c$)
will be corrected from the Newtonian field, which is constructed in
an action-at-a-distance manner, only very slightly (about $10^{-6}$
-- $10^{-4}$ or somewhat larger if cumulative effects are considered)
by general relativistic effects \textit{including} this light-cone
effect (\citealt{Hwang2008}).

Light-cone effect of radiation fields traversing cosmological
distances has been usually approximated by a uniform, spatially-averaged
quantity in semi-analytical studies (e.g. \citealt{Haiman2000} for
LW and H-ionizing backgrounds), or calculated by computationally
expensive, brute-force methods of ray-tracing with finite speed of light
(e.g. \citealt{Bryan2014} for H-ionizing background). The first account
of the inhomogeneous LW background in numerical simulations was taken
by \citet{Ahn2009}, whose consistency with semi-analytical and semi-numerical
approach was confirmed elsewhere (\citealt{Dijkstra2008,Holzbauer2012}).
The method of \citet{Ahn2009} was later used in simulating the process
of cosmic reionization under the influence of the first stars (\citealt{Ahn2012,Fialkov2013}).
This method has been
improved for faster computation using FFT. Its application
to other radiation fields affected by the light-cone effect is 
straightforward, as long as the field strength from the source is
only a function of distance but not of direction. Even though this
improved method was already used in the simulation by \citet{Ahn2012}, its
formalism and generality 
have never been described explicitly. 
Thus we lay out the generic formalism,
a numerical scheme and the actual application in this paper. A similar
account for X-ray background (\citealt{Xu2014}) used
the same machinery of \citet{Ahn2009} with modifications to implement
the opacity due to hydrogen and helium atoms.

This paper is organized as follows. In Section \ref{sec:retarded_field},
we describe the generic formalism and the numerical scheme.
In Section \ref{sec:Applications}, we show how the LW background is calculated
using this scheme, and also quantify the possible error from ignoring the
light-cone effect. In Section \ref{sec:Conclusion}, we conclude our findings and discuss further
prospects and limitations.

\section{Retarded Radiation Field: Formalism\\ and Numerical Method\label{sec:retarded_field}}

\subsection{Green's Function Formalism\label{sub:green}}

A generic radiation field $F({\bf x},\, t)$ at position ${\bf x}$
and time $t$ in a homogeneous and isotropic background spacetime satisfies
a d'Alembertian-type equation. In the trivial case of a Minkowski
background, electromagnetic (EM) waves (see \citealt{Jackson1998})
follow
\begin{equation}
\left(\nabla_{{\bf x}}^{2}-\frac{1}{c^{2}}\frac{\partial^{2}}{\partial t^{2}}\right)F({\bf x},\, t)=-4\pi f({\bf x},\, t),\label{eq:dAlembertian}
\end{equation}
where $F({\bf x},\, t)$ is either the scalar potential $\Phi$ or
the vector potential ${\bf A}$, and $f({\bf x},\, t)$ is the corresponding
source term (the charge density or the current density, respectively).
In this case, $F({\bf x},\, t)$ can be cast into an convoluted integral
form
\begin{equation}
F({\bf x},\, t)=\int d^{3}x'\, dt'\, G({\bf x},\, t;\,{\bf x}',\, t')f({\bf x}',\, t'),\label{eq:integral}
\end{equation}
where the retarded Green's function%
\footnote{We take the simple stance that the initial condition is well established
in astrophysical problems, and therefore the retarded Green's function
is more adequate than the advanced Green's function.%
} (RGF henceforth) $G({\bf x},\, t;\,{\bf x}',\, t')$ satisfies
\begin{equation}
\left(\nabla_{{\bf x}}^{2}-\frac{1}{c^{2}}\frac{\partial^{2}}{\partial t^{2}}\right)G({\bf x},\, t;\,{\bf x}',\, t')=-4\pi\delta({\bf x}-{\bf x}')\delta(t-t')\label{eq:green_diff}
\end{equation}
 and is given in a closed form as
\begin{equation}
G({\bf x},\, t;\,{\bf x}',\, t')=\frac{\delta\left(t'-\left[t-\frac{\left|{\bf x}-{\bf x}'\right|}{c}\right]\right)}{\left|{\bf x}-{\bf x}'\right|}.\label{eq:green_anal}
\end{equation}
Note that Equation (\ref{eq:integral}) is in the most generic form.
The light-cone effect or causality is apparent in equations (\ref{eq:integral})
and (\ref{eq:green_anal}), as the field at observing time $t$ is
caused by a source along the light cone at past time $t'=t-\left|{\bf x}-{\bf x}'\right|/c$.
Even though equations (\ref{eq:dAlembertian}), (\ref{eq:green_diff})
and (\ref{eq:green_anal}) are applicable both to $\Phi$ and ${\bf A}$
only in the Lorentz gauge, it is well known that the electromagnetic
field, as a physical quantity derived from both $\Phi$ and ${\bf A}$,
still conserves this causality under any gauge choice
(\citealt{Jackson2002}; see also a nice
example of Problem 6.20 in \citealt{Jackson1998}). 

Let us consider a generic field $F({\bf x},\,\tau)$ in the spacetime
described by the Robertson-Walker metric with null curvature,
\begin{equation}
ds^{2}=c^{2}dt^{2}-a^{2}(t)d{\bf x}^{2}=a(\tau)\left(d\tau^{2}-d{\bf x}^{2}\right),\label{eq:RW_metric}
\end{equation}
where ${\bf x}$ is the comoving spatial coordinate, $a(t)$ is the
scale factor at conformal cosmic time $\tau$ with normalization $a=1/(1+z)$,
and $\tau$ is defined by 
\begin{equation}
\tau(t)\equiv\int^{t}\frac{c\, dt'}{a(t')}.\label{eq:t_conformal}
\end{equation}
It is convenient to use $\tau$ and ${\bf x}$ because the null geodesic
is simply given by $d\left|{\bf x}\right|/d\tau$=1. One can then
expect that the generic RGF will take the following form: 
\begin{eqnarray}
G({\bf x},\,\tau;\,{\bf x}',\,\tau')&=&\delta\left(\tau'-\left[\tau-\left|{\bf x}-{\bf x}'\right|\right]\right)\nonumber\\
&&\times \mathcal{F}(\left|{\bf x}-{\bf x}'\right|,\,\tau,\,\tau'),\label{eq:generic_green} 
\end{eqnarray}
where $\mathcal{F}(\left|{\bf x}-{\bf x}'\right|,\,\tau,\,\tau')$ is a generic,
geometrical dilution factor. Note that $\mathcal{F}(\left|{\bf x}-{\bf x}'\right|,\,\tau,\,\tau')$,
in general, reflects cosmological effects such as photon-redshifting,
and therefore depends also on $\tau$ and $\tau'$. The corresponding
integral form will become (c.f. Equation \ref{eq:integral})
\begin{equation}
F({\bf x},\,\tau)=\int d^{3}x'\, d\tau'\, G({\bf x},\,\tau;\,{\bf x}',\,\tau')f({\bf x}',\,\tau'),\label{eq:generic_integral}
\end{equation}
where $f({\bf x}',\,\tau')$ is the source term properly defined in
the $({\bf x},\,\tau)$ coordinate system. 

We now decompose Equation (\ref{eq:generic_integral}) into discrete
terms which are affected by different time slices, and also apply
a periodic boundary condition to assimilate the usual simulation set-up.
Equation (\ref{eq:generic_integral}) can be rewritten as
\begin{eqnarray}
&&F({\bf x},\,\tau)\nonumber \\
&&  =  \sum_{n}\int d^{3}x'\,\int_{\tau_{n}}^{\tau_{n+1}}d\tau'\, G({\bf x},\,\tau;\,{\bf x}',\,\tau')f({\bf x}',\,\tau')\nonumber \\
&&  =  \sum_{n}\int d^{3}x'\,\int_{\tau_{n}}^{\tau_{n+1}}d\tau'\,\delta\left(\tau'-\left[\tau-\left|{\bf x}-{\bf x}'\right|\right]\right)\nonumber \\
&&\,\,\,\,\,\,\,\,\,\,\,\, \times \mathcal{F}(\left|{\bf x}-{\bf x}'\right|,\,\tau,\,\tau')f({\bf x}',\,\tau')\nonumber \\
&&  \simeq  \sum_{n}\int d^{3}x'\, W\left(\frac{\left|{\bf x}-{\bf x}'\right|-(\tau-\tau_{n+1})}{\tau_{n+1}-\tau_{n}}\right)\nonumber \\
&&\,\,\,\,\,\,\,\,\,\,\,\, \times \mathcal{F}(\left|{\bf x}-{\bf x}'\right|,\,\tau,\,\tau')f({\bf x}',\,\tau_{n})\nonumber \\
&&  \equiv  \sum_{n}\int d^{3}x'\,\mathcal{F}_{n}(\left|{\bf x}-{\bf x}'\right|;\,\tau)f({\bf x}',\,\tau_{n}),\nonumber \\
&&  \equiv  \sum_{n}F_{n}({\bf x},\,\tau)\label{eq:F_sum}
\end{eqnarray}
where $f({\bf x}',\,\tau_{n})$ approximates the source term during
$\tau'=\left[\tau_{n},\,\tau_{n+1}\right]$ as a fixed 3D quantity,
$W(x)$ is a top-hat window function such that
\[
W(x)=\begin{cases}
1 & {\rm if\,0\le x\le1}\\
0 & {\rm otherwise},
\end{cases}
\]
and the RGF and the partial contribution to $F({\bf x},\,\tau)$ from
the sources during $\tau'=\left[\tau_{n},\,\tau_{n+1}\right]$ are
defined as $\mathcal{F}_{n}(\left|{\bf x}-{\bf x}'\right|;\,\tau)$
and $F_{n}({\bf x},\,\tau)$, respectively. It is important to choose
small enough time gaps such that $f({\bf x}',\,\tau')$ does not evolve
much during each time step. Now, we apply the periodic boundary condition:
the cubical domain with size $L$ (volume $V\equiv L^{3}$) is infinitely
repeated side to side and thus satisfies
\begin{eqnarray}
&& F_{n}({\bf x},\,\tau)  \nonumber \\
&& = \sum_{u,\, v,\, w}\int^{V}d^{3}x'\,\mathcal{F}_{n}\left[\left|{\bf x}-\left\{ {\bf x}'+(u,\, v,\, w)L\right\} \right|;\,\tau\right]\nonumber \\
&& \,\,\,\,\,\,\,\,\,\,\,\,\times f({\bf x}'+(u,\, v,\, w)L,\,\tau_{n})\nonumber \\
&& =  \int^{V}d^{3}x'\,\sum_{u,\, v,\, w}\mathcal{F}_{n}\left[\left|{\bf x}-\left\{ {\bf x}'+(u,\, v,\, w)L\right\} \right|;\,\tau\right]\nonumber \\
&& \,\,\,\,\,\,\,\,\,\,\,\,\times f({\bf x}',\,\tau_{n})\nonumber \\
&&  \equiv  \int^{V}d^{3}x'\,\mathcal{G}_{n}({\bf x}-{\bf x}')f({\bf x}',\,\tau_{n}),\label{eq:3D_integral}
\end{eqnarray}
where $u$, $v$, $w$ are integer indices running from $-\infty$
to $\infty$ along the three axes of the cubical domain, both ${\bf x}$
and ${\bf x}'$ are confined to one domain, the second identity is
guaranteed by the periodicity of the source term, and the last identity
defines the ``effective'' RGF $\mathcal{G}_{n}({\bf x}-{\bf x}')$.
Note that for a given observing point $({\bf x},\,\tau)$ and the
past time-slice at $[\tau_{n},\,\tau_{n+1}]$, the $(u,\, v,\, w)$
points contributing to $\mathcal{G}_{n}({\bf x}-{\bf x}')$ are only
those confined in the spherical shell bounded by comoving radii $\tau-\tau_{n+1}$
and $\tau-\tau_{n}$.

We have just cast the full integral (Equation \ref{eq:generic_integral})
for the retarded field into the summation (Equation \ref{eq:F_sum})
of spatially convoluted integrals (Equation \ref{eq:3D_integral}).
In the next section, we will describe how we calculate the full retarded
field under a uniform Eulerian grid.

\subsection{Numerical Method\label{sub:numerical}}

\begin{figure*}[t!]
\includegraphics[width=1\textwidth]{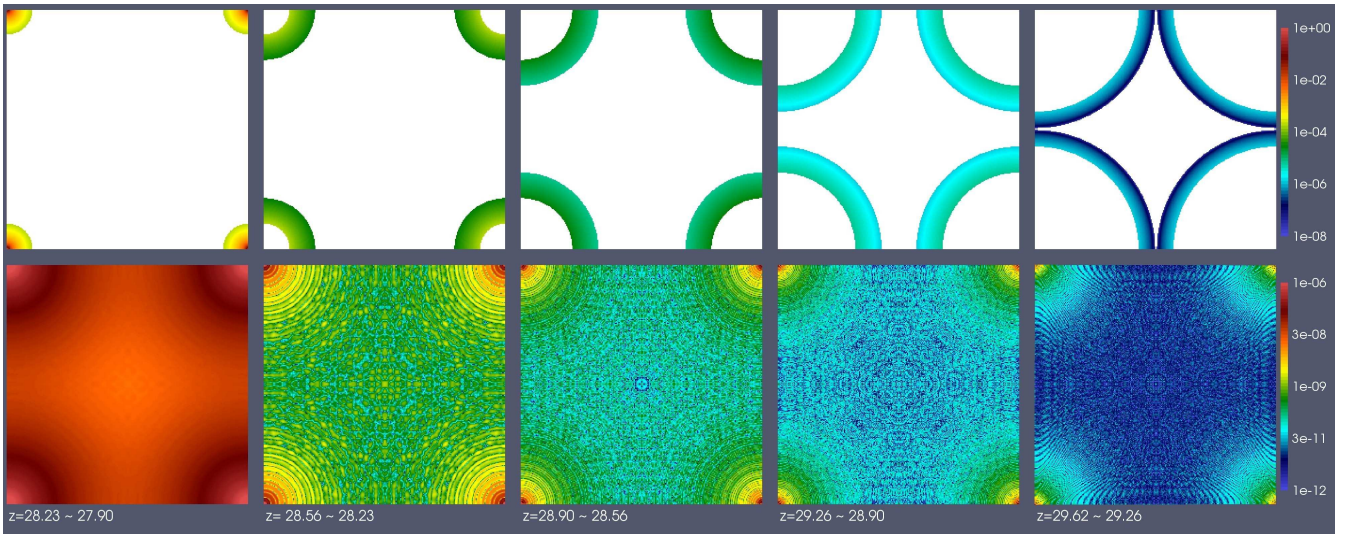}
\caption{RGFs for LW intensity at observing redshift $z=27.90$, in a periodic
box of size $114/h$ (comoving) Mpc and a uniform grid of $256^{3}$
cells. Both the real-space (top row) and Fourier-space (bottom row)
RGFs ($\mathcal{G}_{n}$ and $\hat{\mathcal{G}_{n}}$ respectively)
are shown in log scales, on a slice containing 4 corners of the box.
For $\hat{\mathcal{G}_{n}}$ which has large dynamic range, we took
the absolute value of its real part, $\left|{\rm Re(\hat{\mathcal{G}_{n}})}\right|$,
for better visualization. From left to right, RGFs of progressively
longer lookback-time slices are shown, in units arbitrarily chosen
but consistent throughout time slices. Lookback-time range is shown
in terms of redshift, below each set of $\mathcal{G}_{n}$ and $\hat{\mathcal{G}}_{n}$.
Each redshift range corresponds to 2 Myrs of cosmic time. Note that
the right-most panels correspond to the time slice containing the
horizon $r_{{\rm LW}}$. \label{fig:RGFs-for-LW}}
\end{figure*}

We now develop a numerical method to calculate the retarded
fields. A scheme for calculating 3D fields in numerical simulations
becomes the most transparent under a uniform grid, and thus we restrict
the description also to the case of a uniform grid. 
This choice is further justified by the fact that
a retarded field originates from sources located far from the observing
point, whose impact on each observing point is greatly
averaged out and thus can be accurately calculated even on a uniform grid.

If we discretize relevant quantities on a uniform grid, Equation (\ref{eq:3D_integral})
turns into
\begin{eqnarray}
F_{n}(i,\, j,\,k)&=&\sum_{i'=0}^{N-1}\sum_{j'=0}^{N-1}\sum_{k'=0}^{N-1}\mathcal{G}_{n}(i-i',\,j-j',\,k-k')\nonumber \\
&&\times f_{n}(i',\, j',\, k'),\label{eq:convolution_sum}
\end{eqnarray}
where the domain is divided into cells with size $l\equiv L/N$, $\left\{ i,\, j,\, k\right\} $
and $\left\{ i',\, j',\, k'\right\} $ are integer indices for real-space
positions, and we absorbed $l^{3}$ into $f_{n}(i',\, j',\, k')$
by letting $\int d^{3} x' f({\bf x}',\,\tau_{n}) \longmapsto 
\sum_{i',\, j',\, k'} f_{n}(i',\, j',\, k')$.
If $\hat{A}(p,\, q,\, r)$ is the Fourier component of some field
$A(i,\, j,\, k)$ given by
\begin{eqnarray}
A(i,\,j,\,k)&=&\frac{1}{N^{3}}\sum_{p=0}^{N-1}\sum_{q=0}^{N-1}\sum_{r=0}^{N-1}\hat{A}(p,\,q,\,r)\nonumber \\
&&\times\exp\left[\frac{2\pi {\bf i}(ip+jq+kr)}{N}\right],\label{eq:discrete_FT}
\end{eqnarray}
where the imaginary unit ${\bf i}$ is shown in bold face to avoid
confusion with the index $i$, then the convolution theorem gives
\begin{equation}
\hat{F}_{n}(p,\, q,\, r)=\hat{\mathcal{G}_{n}}(p,\, q,\, r)\,\hat{f}_{n}(p,\, q,\, r),\label{eq:convolution_thm}
\end{equation}
where $\left\{ p,\, q,\, r\right\} $ are the indices of the
Fourier-space coordinates.

The merit of Equations (\ref{eq:convolution_sum}) -- (\ref{eq:convolution_thm})
is that the retarded field can now be calculated by FFT as long as
periodicity is assumed. The algorithm then becomes straightforward,
as follows:
\begin{enumerate}
\item For given (observing) time $\tau$, divide the lookback time into
discrete time slices with $\tau'=[\tau_{n},\,\tau_{n+1}]$.
\item For a given (past) time slice $n$, locate a unit source at the origin
$(i',\, j',\, k')=(0,\,0,\,0)$ of the real-space domain (box), and
attach identical boxes -- each containing one unit source at its origin
-- side by side. The total number of boxes to attach may be limited
by the ``effective horizon'', beyond which the contribution to $\mathcal{G}_{n}(i,\, j,\, k)$
(or to the net $F_{n}(i,\, j,\, k)$ due to a rapid decrease of $f_{n}(i',\, j',\, k')$
in lookback time) becomes too weak to include.
Calculate $\mathcal{G}_{n}(i,\, j,\, k)$ from this configuration
by a direct summation (see its definition in Equation \ref{eq:3D_integral})
over the contributions from these sources, when the functional form
of $\mathcal{F}(\left|{\bf x}-{\bf x}'\right|,\,\tau,\,\tau')$ in Equation (\ref{eq:F_sum})
is known. Note that $\mathcal{G}_{n}(i,\, j,\, k)$ at each observing
point $\left\{ i,\, j,\, k\right\} $ is affected only by the unit
sources inside the spherical shell given by
$\tau-\tau_{n+1}\le r_{\rm os} \le\tau-\tau_{n}$, where $r_{\rm os}$
is the comoving distance between ${\bf x}=(i,\,j,\,k)l$ and ${\bf x}'=(u,\,v,\,w)L$
and the index notation follows those of Equation
(\ref{eq:3D_integral}) and (\ref{eq:convolution_sum}).
\item Get $\hat{\mathcal{G}_{n}}(p,\, q,\, r)$ by applying FFT to $\mathcal{G}_{n}(i,\, j,\, k)$
calculated from step 2.
\item Get $\hat{f}(p,\, q,\, r)$ by applying FFT to $f_{n}(i,\, j,\, k)$.
\item Get $\hat{F}_{n}(p,\, q,\, r)$ from Equation (\ref{eq:convolution_thm}),
and then apply FFT to get $F_{n}(i,\, j,\, k)$.
\item Finally, sum over time slices to get $F(i,\, j,\, k)=\sum_{n}F_{n}(i,\, j,\, k)$.
\end{enumerate}
Note that once the box size, the grid resolution and the output times
are determined, one can precalculate $\mathcal{G}_{n}(i,\, j,\, k)$
for each set of observing time ($\tau$) and lookback time ($\tau_{n}$--$\tau_{n+1}$)
and store all $\hat{\mathcal{G}_{n}}(p,\, q,\, r)$'s regardless of
the source distribution, which is a merit of the convolution theorem.
The number of data files for $\mathcal{G}_{n}(i,\, j,\, k)$ is $\sim N_{{\rm observe}}N_{{\rm lookback}}$,
where $N_{{\rm observe}}$ is the number of observing redshifts and
$N_{{\rm lookback}}$ is the number of lookback times. $N_{{\rm lookback}}$
can depend on the observing redshift, depending on the change of $t_{{\rm lc}}$
over time. When the box size ($L$) is much smaller than the size
of the effective horizon ($r_{h}$), the number of unit sources repeated
around the computation domain will be $N_{{\rm unit-source}}\sim(r_{h}/L)^{3}$,
and the computation of each $\mathcal{G}_{n}(i,\, j,\, k)$ will require
$\sim N_{{\rm mesh}}(r)N_{{\rm unit-source}}$ operations.

\section{Application: Lyman-Werner Background\label{sec:Applications}}

\begin{figure*}[t!]
\includegraphics[width=1\textwidth]{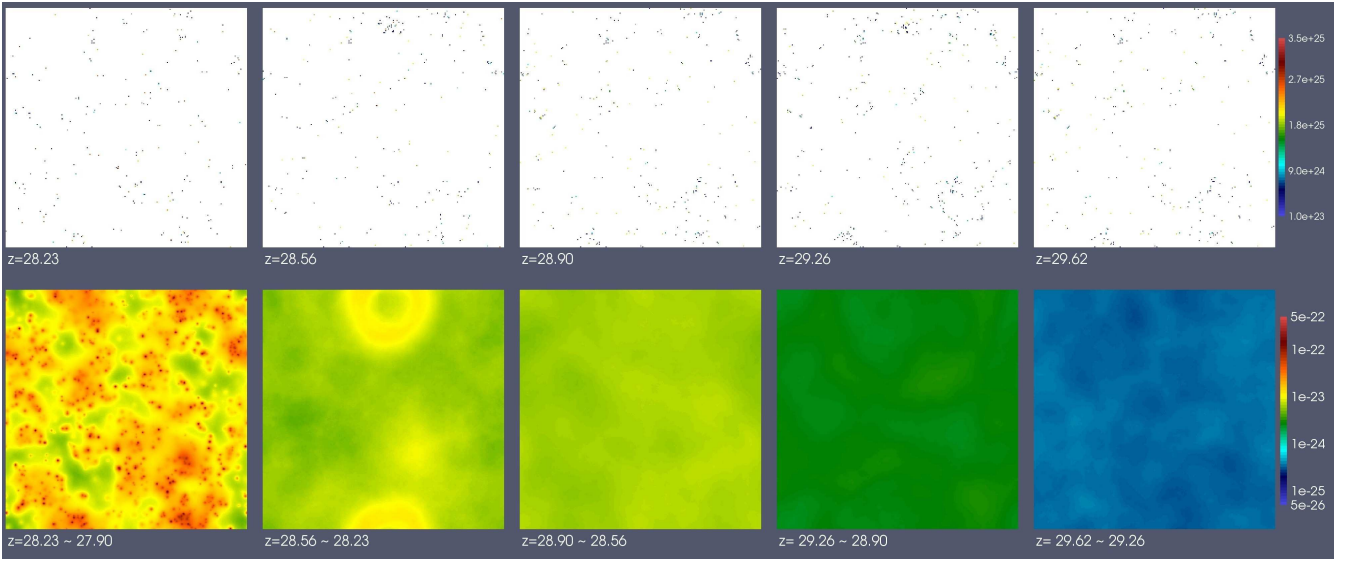}
\caption{Source distribution and LW background from different time slices,
with the same configuration as in Figure \ref{fig:RGFs-for-LW}. The
top row shows the 2D maps of $\left\langle L_{\nu}\right\rangle $
(assumed to be fixed at and beyond the redshift specified below each
panel until the adjacent future redshift) on a slice of thickness
$(114/256)/h$ Mpc, and the bottom row the ``partial'' $J_{{\rm LW}}$'s
on the same slice contributed from the look-back time ranges specified
below each panel. The net $J_{{\rm LW}}$ observed at the observing
redshift $z=27.90$ is given by the summation of all the partial $J_{{\rm LW}}$'s,
which is shown in Figure .\label{fig:JLW_partial}}
\end{figure*}

As an application, we apply the method from Section \ref{sub:numerical}
to the calculation of LW background. LW bands are composed of many
excitation levels of hydrogen molecules (${\rm H}_{2}$), and at high
intensities the LW photons can photo-dissociate ${\rm H}_{2}$. 
Because first stars are born with the help of ${\rm H}_{2}$-cooling,
calculating LW background 
is crucial in the study of first-star formation.

The functional form of the RGF for LW background was first laid out
by \citet{Ahn2009}. The band-averaged and angle-averaged intensity
$J_{{\rm LW}}({\bf x},\, z)$ observed at comoving location ${\bf x}$
and redshift $z$ due to a source at (${\bf x'}$,~$z_{{\rm s}}$)
in units of ${\rm erg}\,{\rm s^{-1}}\,{\rm cm^{-2}}\,{\rm Hz}^{-1}\,{\rm sr^{-1}}$
is given by
\begin{equation}
J_{{\rm LW}}({\bf x},\,
z)=\frac{1}{\left(4\pi\right)^{2}}\frac{\left\langle L_{\nu}({\bf x}',\,z_{s})\right\rangle }{\left|{\bf x}-{\bf x}'\right|^{2}}\frac{(1+z)^{3}}{1+z_{{\rm s}}}f_{{\rm mod}}\left(\left|{\bf x}-{\bf x}'\right|\right),\label{eq:JLW}
\end{equation}
where $\left\langle L_{\nu}({\bf x}',\,z_{s})\right\rangle \equiv\int_{11.5{\rm eV}}^{13.6{\rm eV}}d(h_{p}\nu)L_{\nu}({\bf x}',\,z_{s})/(2.1{\rm eV)}$
is the rest-frame, frequency($\nu$)-averaged source luminosity of
the emitted LW band photons at the source redshift $z_{s}$ and $h_{p}$
is the Planck constant,
\begin{equation}
f_{{\rm mod}}=\begin{cases}
1.7{\rm e}^{-\left(\frac{\left|{\bf x}-{\bf x}'\right|}{116.29\alpha}\right)^{0.68}}-0.7 & {\rm if}\,\frac{\left|{\bf x}-{\bf x}'\right|}{\alpha}\le97.39\\
0 & {\rm otherwise},
\end{cases}\label{eq:fmod}
\end{equation}
and
\begin{equation}
\alpha=\left(\frac{h}{0.7}\right)^{-1}\left(\frac{\Omega_{m}}{0.27}\right)^{-0.5}\left(\frac{1+z_{{\rm s}}}{21}\right)^{-0.5}\label{eq:scaling_LW}
\end{equation}
with the Hubble constant (in units of 100 km/s/Mpc) $h$ and the present
ratio of matter density to the critical density $\Omega_{m}$. Of
course, the light-cone effect is implicit here: $\left|{\bf x}-{\bf x}'\right|=\tau(z)-\tau(z_{{\rm s}})$.
The LW horizon is $r_{{\rm LW}}\equiv 97.39\alpha$, beyond which
no sources contribute to $J_{{\rm LW}}.$ Comparing Equation (\ref{eq:JLW})
to Equations (\ref{eq:generic_green}) and (\ref{eq:generic_integral}), we find that
\begin{eqnarray}
\mathcal{F}(\left|{\bf x}-{\bf x}'\right|,\,\tau,\,\tau')&=& 
\frac{1}{\left(4\pi\right)^{2}\left|{\bf x}-{\bf
      x}'\right|^{2}}\frac{\left[1+z(\tau)\right]^{3}}{1+z_{{\rm
        s}}(\tau')}\nonumber \\
&&\times f_{{\rm mod}}\left(\left|{\bf x}-{\bf x}'\right|\right),\label{eq:F_LW}
\end{eqnarray}
if we take the source term to be a sum of individual source luminosities
such that $f({\bf x}',\,\tau')=\sum_{s}\left\langle
L_{\nu} ({\bf x}_{s}'',\,z_{s}(\tau'))\right\rangle \delta({\bf x}'-{\bf x}_{s}'')$.

\begin{figure*}[t!]
\includegraphics[width=1\textwidth]{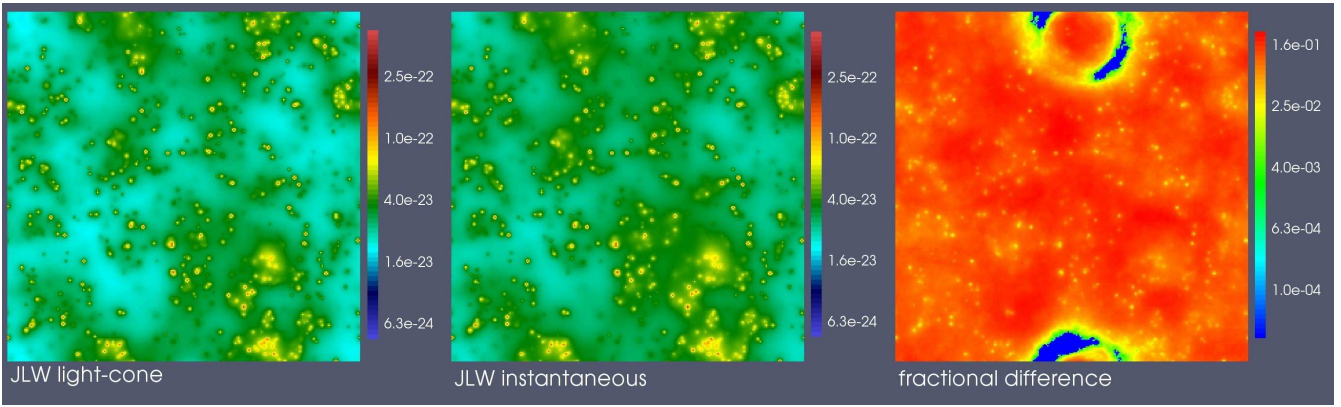}
\caption{Comparison of $J_{{\rm LW}}$'s with the light-cone effect and without.
$J_{{\rm LW}}$ under the correctly implemented light-cone effect
(left), $J_{{\rm LW}}$ under an action at a distance (middle; see
Equation \ref{eq:jlw_instant}) and the fractional difference between
these two fields (right)
are plotted in the same 2D slice as in Figures \ref{fig:RGFs-for-LW}
and \ref{fig:JLW_partial}.\label{fig:JJJ}}
\end{figure*}

When visualized, the real-space RGF $\mathcal{G}_{n}(i,\, j,\, k)$
can give some insight on how $J_{{\rm LW}}$ is constructed. Figure
\ref{fig:RGFs-for-LW} depicts RGFs for $J_{{\rm LW}}$, which clearly
show how the region of influence is determined by the light-cone effect
from a unit source at one corner of the box\footnote{The periodic boundary
condition makes unit sources appear ``around'' every corner of a given
box, as seen in Figure \ref{fig:RGFs-for-LW}.}. Following the
steps described in Section \ref{sub:numerical}, we have calculated
$J_{{\rm LW}}$ and used this to study the formation and suppression
of the first stars during the epoch of reionization (\citealt{Ahn2009,Ahn2012}).
\citet{Ahn2012} simulated the process of cosmic reionization including
the rise and fall of first stars, in a box of size $114/h$ Mpc, on
which Figures \ref{fig:RGFs-for-LW} -- \ref{fig:JJJ} are based.
Lifetime of first stars, which are believed to be as massive as $M_{*}\gtrsim100\, M_{\odot}$
on average (e.g. see a recent result by \citealt{Hirano2014}), is
only about a few Myrs or less, and thus the life and death of these
stars render the source function to change rapidly in time and space.
In Figure \ref{fig:JLW_partial}, the rapid evolution of $\left\langle L_{\nu}\right\rangle $
every 2 Myrs due to first stars is shown, together with the partial
contributions $F_{n}(p,\, q,\, r)$ to $J_{{\rm LW}}$. $r_{{\rm LW}}$
resides within the lookback time of 10 Myrs. One can see that contributions
to $J_{{\rm LW}}$ becomes weaker and more homogeneous as lookback
time increases. Nevertheless, contribution of those time slices with
lookback time $t_{{\rm lookback}}=[2-10]$ Myrs (2nd to 5th panels
from left in Figures \ref{fig:RGFs-for-LW} and \ref{fig:JLW_partial})
to $J_{{\rm LW}}$ is not completely negligible.

How important is the light-cone effect in the LW background from first
stars? To answer this question, we now calculate $J_{{\rm LW}}$ in
an action-at-a-distance way, and then compare this to $J_{{\rm LW}}$
correctly calculated including the light-cone effect. For this, we
now freeze the source distribution at the observing redshift, while
attaching identical boxes side by side for periodicity. Basically,
we now calculate this ``instantaneously affected'' $J_{{\rm LW}}$
as 
\begin{eqnarray}
J_{{\rm LW},\,{\rm instant}}({\bf x},\,z)&=&\frac{1}{\left(4\pi\right)^{2}}\frac{\left\langle L_{\nu}({\bf x}',\,z)\right\rangle }{\left|{\bf x}-{\bf x}'\right|^{2}}\frac{(1+z)^{3}}{1+z_{{\rm s}}} \nonumber \\
&&\times f_{{\rm mod}}\left(\left|{\bf x}-{\bf x}'\right|\right),\label{eq:jlw_instant}
\end{eqnarray}
where $\left\langle L_{\nu}\right\rangle $ is now frozen at the observing
redshift $z$ (e.g. the source term corresponding to the top-leftmost
panel in Figure \ref{fig:JLW_partial}), but we keep all other terms
intact including the implicit condition $\left|{\bf x}-{\bf x}'\right|=\tau(z)-\tau(z_{{\rm s}})$,
in order to quantify only the impact of the temporal evolution of
the source term. Figure \ref{fig:JJJ} shows that the rapid evolution
of source term indeed imprints its signature in $J_{{\rm LW}}$. The
fractional difference $f_{{\rm diff}}\equiv(J_{{\rm LW},\,{\rm instant}}-J_{{\rm LW}})/J_{{\rm LW}}$
is about $\sim10\%$, which is quite significant. Therefore, astrophysical
processes governed by radiation sources of short lifetime should be affected
by the light-cone effect in general.

\section{Discussion\label{sec:Conclusion}}

We have formulated a generic method to numerically calculate radiation fields
which are affected by the light-cone effect. This method
requires discretizing both time and space coordinates, and is analogous
to the usual mesh scheme used in N-body gravity simulations. The light-cone
effect is realized by the retarded-time Greeen's function, and its
calculation becomes efficient when a periodic-box condition
is used. Once the RGFs are calculated, the process of obtaining
a radiation field of interest is performed by FFT, which greatly expedites
computation compared to direct summation over all sources inside
and outside the simulation box.

The relevance of this method is clear in astrophysical problems occurring
on cosmological scales. Once the size of the simulation box is chosen
and the functional form of the RGF is known, one can follow the algorithm
in Section \ref{sub:numerical}, as in the example of LW background
in Section \ref{sec:Applications}. Similar large-scale astrophysical
processes which require an accurate account of the light-cone effect
are the reionization of baryons by UV photons and the reionization
and heating of gas by X-ray photons. It becomes more important at
very high redshift when most astrophysical sources are first stars
and their by-products, because the lifetime of first stars are believed
to be very short and thus the source function evolves  rapidly
in time (Section \ref{sec:Applications}).

There are, of course, limitations to this method. The method is based
on the fact that the RGF is a function only of distance to the source
but not of direction to the source. In reality, radiative transfer is usually affected
by the actual path of the ray in the form of the locally-varying optical
depth. For example, formation of an H II bubble from a source does
not occur in a spherically symmetric way, even though light travels
at the speed of light. This is due to the inhomogeneous distribution
of hydrogen density (and also the optical depth), which then regulates
the H-ionizing background to be dependent on the direction of the
ray. For H-ionizing background, our method can then be used only after
the IGM is fully ionized and the optical depth becomes negligible
everywhere. In case of LW background, we could safely use the method
because the modulation factor $f_{{\rm mod}}$ in Equation (\ref{eq:fmod}),
derived from the cosmic redshifting of LW photons and the frequency
of hydrogen Lyman resonance lines, is valid even when the IGM is ionized,
because simulations of cosmic reionization find that the IGM leaves
trace amount of neutral hydrogen atoms due to recombination (e.g.
\citet{Iliev2008}). In this case, very-high opacity of Lyman resonance
lines guarantees that once LW band photons are redshifted into one
of these lines, even a trace amount of hydrogen atoms can successfully
scatter off those photons, and then $f_{{\rm mod}}$ becomes the only
modulation factor in addition to the usual geometrical dilution factor
(Equations \ref{eq:JLW} and \ref{eq:F_LW}). Another limitation is
that the method cannot treat the gravitational lensing effect, which
depends on the actual mass distribution around each ray's path. If
a problem of interest is found to be strongly affected by the lensing
effect, one should resort back to the ray-tracing method. In addition,
if source locations are resolved beyond the mesh resolution, one can
adaptively choose to use the particle-mesh (PM) scheme to calculate
the near-source influence by direct summation over nearby-source contributions
while calculating far-source influence by our method.

\acknowledgments

We are grateful to the anonymous referee who gave a prompt and
accurate report. This work was supported by the NRF grant funded by
the Korean government MEST (NRF-2012R1A1A1014646; NRF-2014R1A1A2059811).


\end{document}